\pdfoutput=1
\documentclass[aps,pre,twocolumn,superscriptaddress,floatfix,10pt]{revtex4-1}

\usepackage[pdftex]{graphicx} % Allows for eps images
\usepackage{epsfig}
\usepackage{amsmath} % AMS Math Package
\usepackage{amsthm} % Theorem Formatting
\usepackage{amssymb}	% Math symbols such as \mathbb
\usepackage{physics}
\usepackage{float}
\usepackage{xcolor}

\usepackage{hyperref} %clickable references.
\hypersetup{
    colorlinks,
    citecolor=blue,
    filecolor=black,
    linkcolor=red,
    urlcolor=blue
}

\newcommand{\sigds}{{\mathbf{K}_\Delta}}
\newcommand{\muR}{ \boldsymbol{\mu}' }

\newcommand{\Rmu}{ \boldsymbol{A}{\left(r\right)} }

\newcommand{\zv}{\hat{\mathbf{0}}}

\newcommand{\cov}{\mathbf{K}}
\newcommand{\cf}[2]{K{\left(#1, #2\right)}}
\newcommand{\cfR}[2]{K_\Delta{\left(#1, #2\right)}}

\newcommand{\cft}[1]{K{\left(#1\right)}}

\newcommand{\tcor}{\mathbf{K}_2}

\newcommand{\covR}{\mathbf{K}'}
\newcommand{\covij}[2]{K_{#1 #2}}

\newcommand{\covRij}[2]{K'_{#1 #2}}
\newcommand{\sigdsij}[2]{K_{\Delta, #1 #2}}

\newcommand{\bcovij}[2]{\bar{K}_{#1 #2}}
\newcommand{\bcovRij}[2]{\bar{K}'_{#1 #2}}

\newcommand{\muRi}[1]{\mu'_{#1}}

\newcommand{\p}[1]{P{\left(#1\right)}}
\newcommand{\psr}[1]{\p{#1 | R}}

\newcommand{\tp}[2]{P{\left(#1 \to #2\right)}}

\usepackage{soul}

\usepackage[english]{babel}
\usepackage{blindtext}
\usepackage{makecell}
\usepackage{tikz}

\begin{document}

\bibliographystyle{unsrt}

\title{The dynamics of machine-learned ``softness'' in supercooled liquids describe dynamical heterogeneity}
\author{Sean A. Ridout}
\affiliation{Department of Physics and Astronomy, University of Pennsylvania, Philadelphia, PA 19104, USA}
\affiliation{Department of Physics, Emory University, Atlanta, GA 30322, USA}
\author{Andrea J. Liu}
\affiliation{Department of Physics and Astronomy, University of Pennsylvania, Philadelphia, PA 19104, USA}

\begin{abstract}
The dynamics of supercooled liquids slow down and become increasingly heterogeneous as they are cooled. Recently, local structural variables identified using machine learning, such as ``softness," have emerged as predictors of local dynamics. Here we construct a model using softness to describe the structural origins of dynamical heterogeneity in supercooled liquids. In our model, the probability of particles to rearrange is determined by their softness, and each rearrangement induces changes in the softness of nearby particles, describing facilitation. We show how to ensure that these changes respect the underlying time-reversal symmetry of the liquid's dynamics. The model reproduces the salient features of dynamical heterogeneity, and demonstrates how long-ranged dynamical correlations can emerge at long time scales from a relatively short softness correlation length.
\end{abstract}
\maketitle
\section{Introduction}

Sufficiently rapid cooling of a liquid below the melting temperature prevents crystallization and produces a  supercooled liquid. While such a state is not in equilibrium since the true equilibrium state is a crystal, it converges to a well-defined metastable ``equilibrium" state at sufficiently long times~\cite{cavagna2009supercooled}, which obeys detailed balance due to the time-reversibility of the microscopic equations of motion. This state exhibits progressively more dramatic dynamical phenomena with decreasing temperature $T$ with little accompanying change of structure, namely: (1) The time required to reach ``equilibrium,"  the \emph{relaxation time}, grows rapidly with decreasing temperature~\cite{angell1995formation,sastry1998,debenedetti2001}.  In systems known as fragile liquids, this growth is faster than an Arrhenius law. (2) Supercooled liquids are dynamically heterogeneous, with some regions that are relatively mobile (and that tend to relax quickly) and others that are nearly immobile and do not relax.  The strength of this heterogeneity grows as $T$ is decreased~\cite{hetbook,ediger2000review,kob1997dynamical,vidal2000}. (3) At sufficiently low temperatures, the system cannot reach ``equilibrium" and falls into the glass state, where the properties depend on the time since the system was quenched into the glass (the \emph{age} of the glass)~\cite{ediger1996,angell2000,kob1997aging,kob2000,rottler2005,schoenholz2017relationship}.

The super-Arrhenius growth of relaxation time in fragile glassformers suggests that a characteristic energy barrier $\Delta E$ grows as $T$ is lowered ~\cite{dyre2006}. Such an energy barrier should be structural in origin. Moreover, isoconfigurational ensemble simulations demonstrate that at least part of dynamical heterogeneity is directly caused by structural heterogeneity~\cite{widmer2004reproducible}. Nonetheless, characterizing these structural effects has posed a longstanding challenge~\cite{gilman1975,jack2014,chaudhari1979,royall2008}. To identify a useful description of structure, we follow a line of research that uses machine learning to obtain a scalar variable, the softness $S$,  that characterizes the local structure environment around each constituent particle~\cite{cubuk2015identifying,cubuk2016structural,schoenholz2016structural,schoenholz2017relationship,cubuk2020,rocks2021,ridout2022,tah2022kinetic}. Softness characterizes the local structural environment around the particle and correlates strongly with the probability that the particle will soon rearrange. In ``equilibrium" supercooled liquids~\cite{schoenholz2016structural} and aging glasses~\cite{schoenholz2017relationship}, the probability that a particle of softness $S$ will rearrange has an Arrhenius temperature dependence, implying a well-defined energy barrier to rearrangement, $\Delta E(S)$. Here we build a phenomenological theory of the dynamics of supercooled liquids to predict relaxation time and dynamical heterogeneity as functions of temperature.

Past work~\cite{schoenholz2016structural,trap} has proposed simple descriptions of the dynamics of $S$ in thermal systems.  In particular, a softness-based version of the trap model of supercooled liquids, which describe dynamics in terms of non-interacting rearrangements of single particles, has been developed~\cite{trap}. Such a description, however, does not account for anything resembling a well-known effect known as \emph{facilitation}, in which rearrangements of one particle influence the subsequent rearrangements of other particles~\cite{chandler2010}.  There are potentially two ways in which a rearrangement can influence subsequent rearrangements. (1) Through elasticity. A rearrangement gives rise to a strain field that can affect whether rearrangements elsewhere occur~\cite{lemaitre2014}; such an effect is a long-ranged form of facilitation and is typically studied in Elasto-Plasticity (EP) models~\cite{nicolas2018RMP}.  Far-field facilitation due to elasticity has been established as significant at temperatures below the mode-coupling temperature~\cite{rahulMCT}. (2) Through changing local structure. A rearrangement inevitably alters structure nearby; this structural effect is a short-ranged form of facilitation. Both far-field (1) and near-field (2) facilitation have been incorporated in softness-based models of plasticity in athermal systems~\cite{Zhang2021,Zhang2022,Xiao2023}, called Structuro-Elasto-Plasticity (StEP) models. However, because the models are athermal they do not need to obey detailed balance, and they do not.

To understand the importance of detailed balance in quiescent supercooled liquids, consider a simple picture for dynamical facilitation. Local rearrangements can trigger other rearrangements, leading to an avalanche~\cite{candelier2008,candelier2010}, and increasing a particle's softness corresponds to increasing the probability that it will rearrange~\cite{schoenholz2016structural}. Thus one might imagine that a rearrangement might increase $S$ for neighboring particles, thereby facilitating further rearrangements; this effect has been quantified in avalanches in athermal systems under shear~\cite{Xiao2023}.  

In the ``equilibrium" state of a supercooled liquid, however, such a mean change in $S$ is forbidden by time-reversal symmetry. A rearrangement at $t=0$ represents a transition between two locally stable states that are structurally distinct; if properly defined, it remains a rearrangement at $t=0$ upon reversal of time.  Furthermore, although the position of a neighbor may change during the rearrangement, in a continuum or lattice model where $S$ takes a value at some separation $\mathbf{r}$ from the rearrangement, that site is still a distance $\mathbf{r}$ from the rearrangement in the time-reversed case. Measurements of distributions of softness changes due to a rearrangement, $\Delta S{\left(r\right)}$, must be fit to forms consistent with this time-reversal symmetry.  

Including detailed balance in EP or StEP models is difficult because they have no Hamiltonian~\cite{nicolas2018RMP}. 

Here we neglect elasticity and far-field facilitation, developing a theoretical framework for the spatiotemporal evolution of softness that allows for near-field facilitation and yet obeys detailed balance.  The resulting Time-Reversible Structuro-Plastic (TRSP) models should be useful above the mode-coupling temperature.  The general approach could be used for \emph{any} measure of the local structural environment~\cite{richard2021} that satisfies certain properties, not just softness. For such models based on softness, we identify the few model parameters using direct measurements of the distribution of $S$, the probability of particles of a given $S$ to rearrange, and the distribution of $S$ changes induced by rearrangements in MD simulations of a standard supercooled liquid model. We demonstrate that one of these models semi-quantitatively reproduces many features of relaxation and dynamical heterogeneity, and show how large observed dynamical correlation lengths originate from a small softness correlation length within the model.

\section{MD simulation model and computation of softness}

We study the standard Kob-Andersen Lennard-Jones mixture, of 80\% large and 20\% small particles. We simulate systems of $N=10,000$ particles. After equilibrating at a given temperature $T$, we study the dynamics using NVE simulations. 

As in past work~\cite{schoenholz2016structural}, we define the softness $S$ by training a classifier to distinguish particles that undergo large rearrangements from those which do not rearrange for a long time using local structural variables. We identify rearrangements using the rearrangement indicator $p_{\mathrm{hop}}$, defined as as in~\cite{schoenholz2016structural}.  We construct a training set at $T=0.47$ using rearrangements with $p_{\mathrm{hop}} > 0.6$ as our rearranging examples and particles which have $p_{\mathrm{hop}} < 0.02$ for $1000\tau \approx 2 \tau_\alpha$ as the non-rearranging examples.  In later analyses we consider all particles with $p_{\mathrm{hop}} > 0.2$ to be rearranging, as justified in~\cite{schoenholz2016structural}.

We then train a SVM classifier to distinguish our rearranging and non-rearranging examples using $M=266$ structure functions $g_{\alpha,i}$, similar to those used in~\cite{cubuk2015identifying}. These contain a combination of radial structure functions which measure the density of neighbours of particle $i$ and angular structure functions which characterize the angles between triplets of particles. Details are specified in Appendix \ref{app:sf}.  This produces weights $\mathbf{w}$ such that the sign of the softness

\begin{equation}
S_i = w_0 + \sum_{\alpha} w_\alpha g_{\alpha,i}
\end{equation}
is maximally predictive of whether particle $i$ is rearranging or not.

For the MD simulations, we report lengths and energies in units of the large-large  interaction distance $\sigma_{AA}$ and scale $\epsilon_{AA}$, and times in units of $\sqrt{m \sigma_{AA}^2 / \epsilon_{AA}}$.

As in past work~\cite{schoenholz2016structural,schoenholz2017relationship,trap} we consider only the large particles in our analyses.

\section{Enforcing time-reversal symmetry: building a TRSP model}

Our aim is to build a model for the spatiotemporal evolution of softness $S$ directly from quantities measured in MD data, \textit{i.e.} the distribution of $S$, its spatial correlations, its effect on rearrangement rates, and how it changes due to a rearrangement. We follow the tradition of elasto-plastic models~\cite{nicolas2018RMP} and implement our model on a lattice, with its state described by a vector $\mathbf{S}$, where the vector component $S(\vec r)$ denotes the softness of the site at $\vec r=r \hat r$.

The softness distribution is approximately Gaussian and $S$ has short-ranged, exponentially decaying spatial correlations $C_S(r)$~\cite{schoenholz2016structural, cubuk2016structural}.  We thus want to ensure that, at equilibrium, $\mathbf{S}$ is drawn from a multivariate normal distribution, with a covariance matrix $\cov$: 

\begin{equation}
P{\left(\mathbf{S}\right)} \propto e^{-\left(\mathbf{S} - \mu \right) \cov^{-1} \left(\mathbf{S} - \mu \right)^T }.
\end{equation}

Note that each of the diagonal terms of the covariance matrix, $\covij{i}{i}$, has the same value, $\sigma^2$, and the off-diagonal terms depend only on the distance $r_{ij}$ between sites $i$ and $j$ so that $\covij{i}{j}= \sigma^2 C_S(r_{ij})$.
In our lattice model, we approximate the dynamics of the particulate system with a sequence of discrete ``rearrangements.'' These are detected in the MD simulation as events when some indicator of single-particle motion (in our case, $p_{\mathrm{hop}}$, computed on inherent-state positions as in~\cite{schoenholz2016structural}), rises above some threshold. The probability of a particle rearranging is found, in simulation data, to depend on its softness through the Arrhenius law

\begin{equation}
P{\left(R|S\right)} \propto e^{\Sigma(S) - \Delta E(S)/T }.
\end{equation}

As in ~\cite{schoenholz2016structural,trap}, $\Sigma(S)$ and $\Delta E(S)$ are roughly linear in $S$, and deviations can be adequately treated by quadratic corrections~\cite{trap} (see Fig. \ref{fig:prs}). Thus, we also have

\begin{equation}
P{\left(R|S\right)} = A{\left(T\right)} e^{\gamma{\left(T\right)} S - \delta{\left(T\right)} S^2},
\end{equation}
with $\gamma(T)$ and $\delta(T)$ related to the linear and quadratic coefficients in $\Sigma(S)$ and $\Delta E(S)$. We assume continuous time dynamics consisting of a series of instantaneous ``rearrangements'' that occur with a rate proportional to the $P(R|S)$ measured in MD simulations.

To formulate precisely how detailed balance constrains the distribution of $\Delta S$ for a rearrangement, we construct the distribution of $\mathbf{S}$ given that a site at the origin is rearranging, using Bayes' theorem:

\begin{equation}
    P{\left(\mathbf{S} | R\right)} \propto P{\left(R | S{\left(0\right)} \right)} P{\left(\mathbf{S}\right)}.
\end{equation}

Clearly, $P{\left(\mathbf{S} | R\right)}$ must be a normal distribution whose mean and covariance we denote by $\muR$ and $\covR$.  Direct calculation (Appendix \ref{app:fulldb}) shows that $\muRi{i}$ at a distance $r_i$ from the rearranging site, and $\covRij{i}{j}$ for sites $i$ and $j$ at distances $r_i$ and $r_j$ from the rearranging site and with separation $r_{ij}$, are given by

\begin{align}
    \covRij{i}{j} &= \covij{i}{j} -\frac{ 2 \delta  \covij{0}{i} \covij{0}{j}}{1 + 2 \delta \covij{0}{0}} \label{eq:covR} \\ &= \sigma^2 \left(C_S(r_{ij}) -\frac{ 2 \delta    C_S(r_i) C_S(r_j) }{1 + 2 \sigma^2 \delta }\right) \\
   \muRi{i} &= \mu + \left(  \gamma - 2\delta \mu \right) \covRij{0}{i} \\
&= \mu + \frac{\gamma - 2 \delta \mu}{1 + 2 \delta \sigma^2} \sigma^2 C_S(r_i) \label{eq:meanR}
\end{align}

Now consider the vector $\Delta \mathbf{S}$, the change of softness induced by a rearrangement at the origin. We approximate the distribution of $\Delta \mathbf{S}$ as Gaussian as well, with mean $\langle \Delta \mathbf{S} \rangle$ and covariance matrix $\sigds$.   Fig. \ref{fig:dS}(b) shows that $\langle \Delta S_i \rangle$ is roughly a linear function of $S_i$. The covariance $\sigds$ for two sites $i, j$ should decay both as they become separated from each other as well as from the rearrangement at the origin, since the effect of the rearrangement should be short-ranged.  We find from MD simulations that after subtracting a background we associate with undetected rearrangements, $\sigdsij{i}{i}$ appears to be independent of $S_i$ (Fig. \ref{fig:dS}(a)). Thus, we assume that $\sigds$ is independent of $\mathbf{S}$. The simplest possible model therefore has the form

\begin{equation}
\langle \Delta \mathbf{S} \rangle = \mathbf{H} \left( \mathbf{A} - \mathbf{S} \right). \label{eq:linear}
\end{equation}

If the matrix $\mathbf{H}$ is taken to be diagonal, this means that a rearrangement tends to restore a site's softness toward some ``target'' value $A{\left(r\right)}$ with some ``strength'' $ \eta{\left(r_i\right)} \equiv H_{ii}$; if $\mathbf{H}$ has off-diagonal elements then the softness of one's neighbor influences this mean change.

Under these assumptions, the joint distribution of the softness field before and after a rearrangement, $\left(\mathbf{S}, \mathbf{S}' = \mathbf{S} + \Delta \mathbf{S} \right)$, is Gaussian. This may be seen by explicitly writing out the density

\begin{equation}
P{\left(\mathbf{S}, \mathbf{S}' | R\right)} \propto P{\left(\mathbf{S}\right)} P{\left(R | S(0)\right)} P{\left(\Delta \mathbf{S} | R, \mathbf{S}\right)}
\end{equation}
using our assumed forms for all three probabilities.

In order for the system to satisfy detailed balance, the joint distributions of $\mathbf{S}$ for all particles before and after the rearrangement must be equal:
\begin{equation}P{\left(\mathbf{S}| R\right)} = P{\left(\mathbf{S}' | R\right)}.
\label{eq:detbal}
\end{equation}

Applying this constraint to a rearrangement at the origin places conditions on the distribution of $\Delta \mathbf{S}$: a generic distribution of $\Delta \mathbf{S}$ will not be consistent with detailed balance between the states $\mathbf{S}$ and $\mathbf{S}'=\mathbf{S}+\Delta \mathbf{S}$.  Instead of working with Eq.~\ref{eq:detbal} directly, it is far easier to derive these conditions by using the fact that, as stated above, $\mathbf{S}$ and $\mathbf{S}'$ have a joint Gaussian distribution in the case of a linear correlation between $\Delta \mathbf{S}$ and $\mathbf{S}$. We construct the block covariance matrix of the random variable $\tilde{\mathbf{S}}= \left( \mathbf{S}, \mathbf{S}'\right)$:

\begin{align}
    \tilde{\mathbf{K}} &= \left( \begin{array}{cc} \covR & \tcor \\ \tcor & \covR \end{array}\right).
\end{align}

The fact that the rearrangement obeys time-reversal symmetry captured by the equality of the two diagonal blocks and the equality of the two off-diagonal blocks. Since a linear transformation $Y = A X$ of a Gaussian random variable $X$ with covariance $\Sigma_X$  and mean $\langle X \rangle$ yields $\Sigma_Y  = A \Sigma_X A^T$ and $\langle Y \rangle = A \langle X \rangle$, we may easily construct the covariance and mean of $\mathbf{y} = \left(\mathbf{S}, \Delta \mathbf{S}\right)$

\begin{align}
    \langle \mathbf{y} \rangle &= \left( \muR, \mathbf{0}\right) \\
    \mathbf{K}_y &= \left(\begin{array}{cc}
    \mathbf{K} & \mathbf{K}_2 - \mathbf{K} \\ \mathbf{K}_2  - \mathbf{K} & 2 \left(\mathbf{K} - \mathbf{K}_2\right)
    \end{array} \right).
\end{align}

Finally, we condition on $\mathbf{S}$ to construct $P{\left(\Delta \mathbf{S} | \mathbf{S}\right)}$  (Appendix \ref{app:fulldb}). After identifying $\mathbf{H} =   1\!\!1  - \tcor\covR^{-1}$, this yields the conditions

\begin{align}
    \mathbf{A} &= \muR\label{eq:S0}\\
    \sigds &= \left(2 - \mathbf{H}\right) \mathbf{H} \covR. \label{eq:covar}
\end{align}

To understand the meaning of these conditions, consider a simple model without spatial correlations in $S$ or $\Delta S$. In this case $H_{ij} = \delta_{ij} \eta(r_i)$ and $\sigdsij{i}{j} = \delta_{ij} \sigma^2_{\Delta}(r_{i})$, and the latter condition reduces to

\begin{align}
    \sigma_\Delta^2(r) &= \sigma^2 \eta(r) \left( 2 - \eta(r) \right). \label{eq:nocorr}
\end{align}

Thus, we can see that these conditions state that the variance of the kick, $\sigma^2_{\Delta}(r)$, and the degree to which $S$ is ``restored'' toward the mean $S$, $\eta(r)$, cannot be independent of each other. This is intuitively reasonable from time-reversibility: if the restoring term is much stronger (weaker) than the random kick, then the distribution at distance $r$ will tend to narrow (broaden) due to the rearrangement. But since the rearrangement is still at distance $r$ from the site under consideration in the time-reversed trajectory, this violates time-reversal symmetry.  

If there are spatial correlations in $S$ and $\Delta S$, then detailed balance constrains the relation between the covariance of the kick, $\sigds$, and the degree to which $\mathbf{S}$ is restored to its mean, $\mathbf{H}$ as in Eq.~\ref{eq:covar}. In other words, if we measure the covariances $\cov, \sigds$ of $\mathbf{S}$ and $\Delta \mathbf{S}$ in simulations, detailed balance constrains the value of $\mathbf{H}$.  Assuming the necessary square root exists, we have the unique solution~\cite{quadraticpaper}~\footnote{In general the square root of a matrix is not unique---for each eigenvalue we have a choice about whether to choose the positive square root or the negative square root. However, we must always choose the positive square root; if $\mathbf{H}$ has an eigenvalue greater than $1$, it corresponds to an unphysical solution in which $S$ on average overshoots its equilibrium value.} 

\begin{equation}
\mathbf{H} = 1\!\!1 -  \sqrt{ 1\!\!1  - \sigds \covR^{-1}}. \label{eq:sqrt}
\end{equation}
For the square root to exist, the eigenvalues of $\sigds \covR^{-1}$ must be less than 1; this generalizes an observation in the uncorrelated case that the variance of $\Delta S$ for the rearranging particle must not exceed the variance $\sigma^2$ of the softness itself. 

\section{Generalizing TRSP models to multiple-particle rearrangements}

In the above TRSP model, we have assumed that a single site ``rearranges'' at a time. This model underestimates the size of dynamical heterogeneity (see Appendix~\ref{app:singlesite}). In reality, rearrangements must involve multiple neighbouring particles passing over the threshold $p_c$; this ``rearrangement size'' contributes to the measured dynamical correlations. Note that the rearrangement size is a \emph{same-time} correlation; when a particle rearranges there are usually neighbours rearranging with it. Indeed, if one considers an elementary T1 event, four particles must rearrange in tandem.

We therefore generalize our rearrangement rule based on $P{\left(R|S\right)}$ and our distribution of $\Delta \mathbf{S}$ to allow for rearrangements composed of multiple particles.   For simplicity we assume that all rearrangements involve exactly $m$ particles. Once $m$ is chosen, we consider only rearrangements involving $m$ adjacent particles. We assume that the rate of rearrangement for a cluster of particles is determined by the average $S$ of the cluster, i.e. $P_0{\left(R|\langle S \rangle_m\right)}$.  Furthermore, we assume that the variance of $\Delta S$ for a particle is determined by its distance from the \textit{nearest} rearranging particle.  These choices are the simplest one which preserve the symmetry between the $m$ rearranging particles and other basic physical considerations. In Appendix \ref{app:kcorr} we discuss the problems with other ``obvious'' choices for the multi-particle rearrangement rule.

This change in the model requires a ``renormalization'' of $P{\left(R|S\right)}$.  The observed $P{\left(R|S\right)}$ is now related to the ``bare'' function $P_0{\left(R|\langle S \rangle_m \right)}$ by an average over the neighbour softness $S'$, conditioned on a given particle's $S$. We carry out the necessary Gaussian integral and solve the resulting equations for the parameters of the bare $P_0{\left(R|S\right)}$ in terms of the observed parameters. This calculation, sketched in Appendix \ref{app:renorm}, results in

\begin{widetext}
\begin{align}
\delta_0{\left(T\right)} &=  \frac{\delta}{c^2 - 2 c \left(1 - c\right) \delta \sigma^2}  = \frac{\delta}{c^2} + \order{\delta^2 \sigma^4} \label{eq:delta_renorm} \\
\gamma_0{\left(T\right)} &=\frac{\gamma + 2 c \left(1 - c\right) \delta_0 \left(\mu + \gamma \sigma^2\right)}{c} = \frac{\gamma}{c} + \order{\delta \sigma^2} \label{eq:gamma_renorm}\\
A_0{\left(T\right)} &= \frac{A}{z} \sqrt{\frac{1 + 2 c \delta_0 \sigma^2}{1 + 2 \delta \sigma^2}} \exp(\frac{\gamma \mu - \delta \mu^2 + \gamma^2 \sigma^2/2}{1 + 2 \delta \sigma^2}-\frac{\gamma_0 \mu - \delta_0 \mu^2 + \gamma_0^2 \sigma^2/2}{1 + 2 \delta_0 \sigma^2}), \label{eq:A_renorm}
\end{align}
\end{widetext}
where

\begin{align}
c = \frac{1}{\sigma^2}  \bcovij{R}{R'}
\end{align}
is the average correlation of $S$ between particles $R$, $R'$ in the rearranging cluster, and $z$ is the ratio between the number of clusters $N_m$ and the number of sites $N$.

 The question remains of how to implement the softness change $\Delta \mathbf{S}$ produced by a multi-particle rearrangement in a manner which preserves time-reversal symmetry.  To begin to address this question, we must compute the distribution $P{\left(\mathbf{S}|R_{i_0, i_1, \dots i_m}\right)}$ of $\mathbf{S}$ conditioned on a particular set of rearranging particles $i_0, i_1, \dots, i_m$.  A straightforward calculation generalizing equations \ref{eq:covR}--\ref{eq:meanR} to multi-particle rearrangements (Appendix \ref{app:fulldb}) shows that the covariance matrix and mean are given by

 \begin{align}
\covRij{i}{j} &= \covij{i}{j} - \frac{2 \delta \bcovij{R}{i} \bcovij{R}{i}}{1 + 2 \delta \bcovij{R}{R'}} \label{eq:covmult} \\
\muRi{i} &= \mu + ( \gamma - 2\delta \mu ) \bcovRij{R}{i} \\
&= \mu + \frac{ \gamma - 2 \delta \mu}{1 + 2 \delta \bcovij{R}{R'}} \bcovij{R}{i} \label{eq:meanmult},
 \end{align}
where $\bcovij{R}{i}$ denotes the average of $K_{Ri}$ for every rearranging site $R$ and $\bar{K}_{RR'}$ denotes the average over all pairs of (not necessarily distinct) rearranging sites $R, R'$. 

We may preserve detailed balance by choosing any functional form for $\sigds$ that treats the different rearranging sites $R$ on equal footing, using our previously-derived detailed-balance conditions \ref{eq:covar}, with the new value of $\covRij{i}{j}$ in equation $\ref{eq:sqrt}$, and restoring now toward the newly-derived $\boldsymbol{\mu}$.  Two obvious choices are a $\sigds$ which depends only on the distance to the center of mass of the rearrangement, and our actual choice, a $\sigds$ which depends only on the closest distance to any rearranging site.  

Inspired by T1 events, we will consider the case $m=4$, and specifically restrict rearrangements to clusters of $4$ particles arranged in a nearest-neighbor square (plaquette).

\section{Measuring TRSP model parameters from MD simulation}

The model is determined by $\Delta E(S)$, $\Sigma(S)$, $\cov$, $\covR$, and $\langle S \rangle$, which we extract from simulations. Detailed balance then fixes $H$ and $\Rmu$ at each temperature. 

As in~\cite{trap}, we assume that 

\begin{align}
    \Delta E{\left(S\right)} &= \epsilon_0 + \epsilon_1 S + \epsilon_2 S^2 \label{eq:dE} \\
    \Sigma{\left(S\right)} &= \Sigma_0 - \frac{\epsilon_1}{T_0} S - \frac{\epsilon_2}{T_0} S^2 \label{eq:s}.
\end{align}
Here the parameter $T_0$ is an ``onset temperature'', above which dynamics no longer depend on the structure as measured through $S$~\cite{schoenholz2016structural}. Here we obtain this onset temperature through the best fit of $\Delta E, \Sigma$ to the data and obtain a noticeably lower temperature than in past work ($T_0 \approx 0.74$ rather than $T_0\approx 0.82$). The discrepancy is likely due to the wider range of $S$ considered; at high $S$ the quadratic correction to $P(R|S)$ has a stronger effect. Note that we find this difference in $T_0$ both with and without rate renormalization.

\begin{figure}[h]
    \includegraphics[]{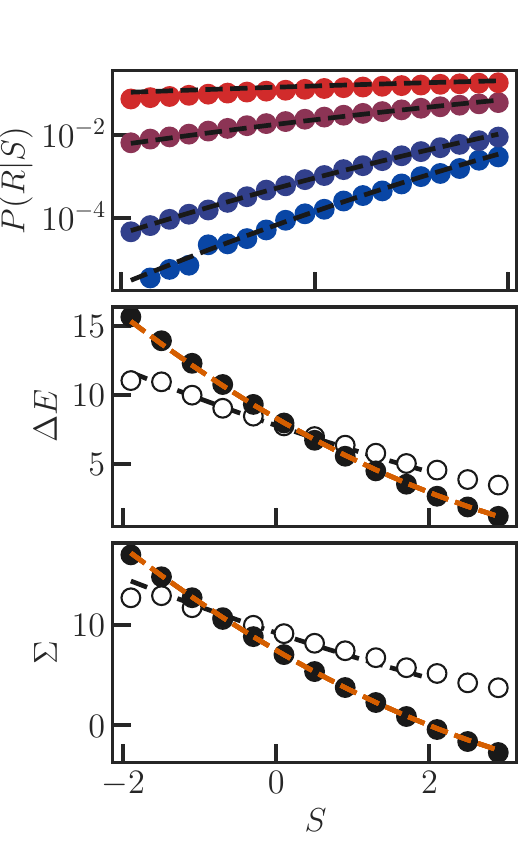}
    \caption{Probability of rearrangement as a function of $S$. (a) $P(R|S)$ at $T=0.7, 0.6, 0.48, 0.43$. Black dashed lines are renormalized fits, as described in the text. (b) Effective energy barrier $\Delta E(S)$, and (c) entropic factor $\Sigma(S)$. In both (b,c), white circles are values inferred directly from $P(R|S)$, while black circles represent ``bare'' values which, when renormalized for the cubic lattice used in the model, give the black dashed lines in (a). Orange dashed lines are fits to a quadratic function of $S$ with an onset temperature $T_0$, eqs. \ref{eq:dE},\ref{eq:s}.}
\label{fig:prs}
\end{figure}

Fig. \ref{fig:prs}(a) shows $P(R|S)$, measured in MD simulations, at several temperatures. Fig. \ref{fig:prs}(b,c) show the fit values of $\Delta E(S)$ and $\Sigma(S)$ (white circles), and the inferred bare values (black circles). The dashed orange lines show fits to \ref{eq:dE} and \ref{eq:s}. The dashed black lines in Fig. \ref{fig:prs}(a) are the result of passing the bare values of $\Delta E(S)$ and $\Sigma(S)$ back through equations \ref{eq:delta_renorm}-\ref{eq:A_renorm}, showing that the renormalized fit describes $P(R|S, T)$ adequately. This confirms our choices that entered into the renormalization.

\begin{figure}[h]
    \includegraphics[]{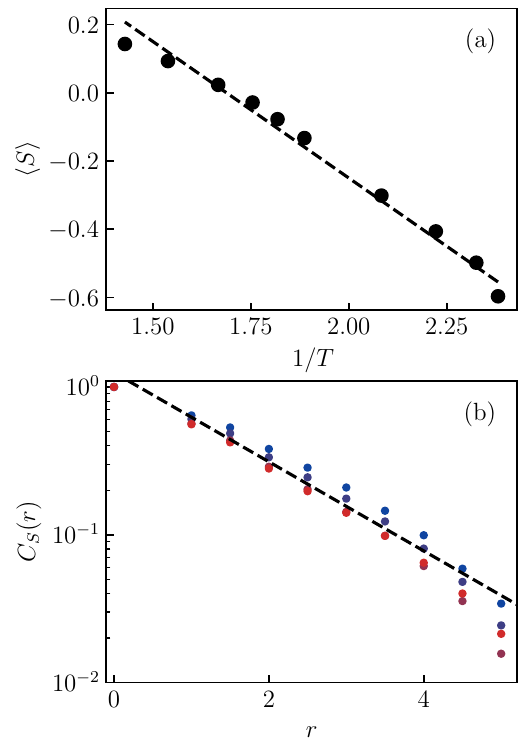}
    \caption{(a) Mean $\langle S \rangle$ vs.~$1/T$ in MD simulations (points), with the fit (dashed) used to interpolate and extrapolate in $T$. (b) Softness spatial correlation function $C_S(r)$ measured from MD simulations (points) at several different temperatures $T$. Colors indicate temperatures listed in caption to Fig.~1(a). Because the $T$-dependence is weak we neglect it and use the fit shown by the black dashed line, Eq. \ref{eq:modexp}.}  
\label{fig:meancorr}
\end{figure}

We must now fit the correlation function $\cov$ and the mean softness $\mu(T)$. (Note that alternatively the model could be parameterized using the $S$ at which the mean $\Delta S$ for rearranging particles is zero at each $T$ instead of $\langle S \rangle$.)  We measure the mean $\langle S \rangle$ (Fig. \ref{fig:meancorr}(a)) and correlation function $C_S(r)$ (Fig. \ref{fig:meancorr}(b) of $S$ as a function of $T$. $\langle S \rangle$ is adequately approximated by an Arrhenius-like form, $a + b/T$. We find that $\cov$ is well-fit by a function of the form

\begin{equation}
C{\left(r\right)} = \begin{cases} C_0 , & r = 0 \\ C_1 e^{-r/\xi}, & r > 0, \label{eq:modexp}\end{cases}
\end{equation}
where $C_0 = \sigma^2$, $C_1  > \sigma^2$.  The correlation length $\xi$ appears to grow slightly with decreasing temperature, as expected, but for now we neglect this effect and fit the correlation function at high $T$.

All that now remains is to measure $\sigds$. In principle $\cfR{\mathbf{r}}{ \mathbf{r'}}$ may be an arbitrary function of $\mathbf{r}$, $\mathbf{r}'$. However, binning the data for rearrangements in $r, r'$, and the angle between them produces data which are too noisy for reliable inference of $\sigds$. To make progress we approximate $\sigds$ by the factorization \begin{equation}
\cfR{\mathbf{r}}{\mathbf{r'}} =  \sigma_\Delta{\left(\tilde{r}\right)} \sigma_\Delta{\left(\tilde{r}'\right)} \rho{\left( \left| \mathbf{r} - \mathbf{r}' \right| \right)}, \label{eq:factorsig}
\end{equation} where $\tilde{r}$ ($\tilde{r}'$) is the \textit{shortest} distance between $\mathbf{r}$ ($\mathbf{r}'$) and a rearranging particle and $\rho{\left(0\right)} = 1$.  

\begin{figure}[h]
    \includegraphics[]{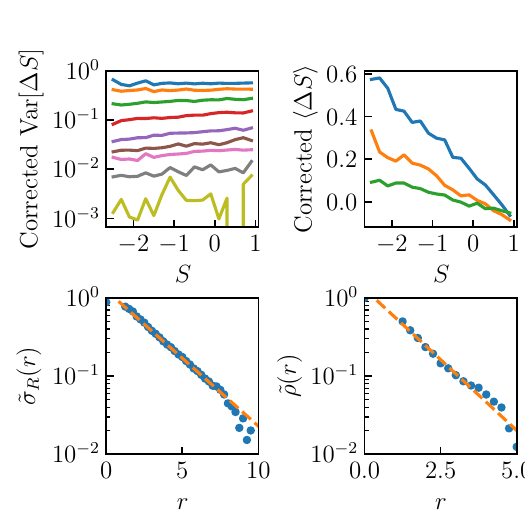}
    \caption{Softness changes $\Delta S$ measured in MD simulations for particles at distance $r$ from the nearest rearrangement, after subtracting $r=\infty$ background as described in Appendix \ref{app:subtract}. (a) Corrected $Var[\Delta S]$ as a function of $S$ for the same particle for $r=1.25, \dots, 7.75$, showing rough independence of $S$. (b) Corrected $\langle \Delta S\rangle$ as a function of $S$ for the same particle for $r=1.25,1.75,2.25$, showing rough linearity in $S$. (c) Corrected standard deviation of $\Delta S$,averaged over $S$, as a function of $r$. (d) Correlation between rearranger's $\Delta S$ and $\Delta S$ for a neighbor at distance $r$. Note that (c,d) are fit to modified exponential decays, as described in the text.}
    \label{fig:dS}
\end{figure}

First we measure $\sigma_\Delta{\left(r\right)}$, i.e. the magnitude of the softness change at distance $r$.  To do so, we look at the variance of $\Delta S$ for particles which are distance $r$ to the closest rearranging particle, over the span of that rearrangement. For each rearrangement we compare $\mathbf{S}$ at times $\delta t = \pm 5 \tau$ relative to the rearrangement. There is a small nonzero plateau of this variance as $r\to \infty$; this plateau value  is larger for particles with large $S$. This is consistent with an interpretation of this plateau as resulting from undetected rearrangements below the threshold $p_c$. We ``subtract off'' this effect of undetected rearrangements, as detailed in Appendix \ref{app:subtract}.  Doing so results in a variance which is roughly independent of $S$ at each $r$ (Fig \ref{fig:dS}(a)), as assumed above.  We then average out the fluctuations over all $S$ bins, producing a numerical estimate $\tilde{\sigma}(r)$ of $\sigma(r)$.  We measure $\rho{\left(\left|\mathbf{r} - \mathbf{r}' \right|\right)}$ by measuring $\tilde{\rho}(r) \equiv \langle \Delta S{\left(0\right)} \Delta S{\left(r\right)} / \tilde{\sigma}(0) \tilde{\sigma}(r)$ for rearrangements at the origin.   

As shown in Fig. \ref{fig:dS}(c,d), we find that $\sigma_\Delta{\left(r\right)}$ and $\rho{\left(r\right)}$ are well fit by the almost-exponential function given by Equation \ref{eq:modexp}. We assume that $\sigds$ is independent of temperature---at high temperatures there may be many more rearrangements, but since a rearrangement is defined as a structural change of a certain magnitude, it is reasonable to assume that the effect of an individual rearrangement is roughly independent of temperature. 

After determining the form of $\sigds$, the form of $\mathbf{H}$ is fixed by time-reversal symmetry through Eq.~\ref{eq:sqrt}, which we evaluate numerically. Note that since we are placing the model on a square lattice, this will in principle produce a slightly anisotropic $\mathbf{H}$.

We find that, at all temperatures, the average duration of a rearrangement, measured by the number of consecutive frames (separated by $\tau$) where $p_{\mathrm{hop}} > 0.2$, given that $p_{\mathrm{hop}} > 0.2$ for at least one frame, is roughly $4\tau$. Thus, the rate of rearrangement is taken to be $P(R|S, T) / 4$.

\section{TRSP model calculations}

After fitting the model parameters, we simulate the model on a cubic lattice of linear dimension $L=20$ with periodic boundary conditions. We take the lattice spacing to be equal to the first peak of the large particle-large particle pair correlation function $g_{AA}(r)$.

We use Gillespie's method to determine each rearrangement event in sequence\cite{Gillespie} rather than using a fixed timestep $\Delta t$.  Multivariate Gaussian random variables are generated using a standard FFT-based technique for translation-invariant covariance matrices~\cite{book}.  

At each temperature $T$ we initialize the system with a state drawn from the equilibrium distribution $\mathcal{N}{\left(\mu{\left(T\right)}, \cov\right)}$. We verify that $\langle S\rangle$ and $C_S(r)$ do not change in time from the putative equilibrium values, as expected from our detailed-balance conditions (Appendix \ref{app:check_mc}). We also check explicitly that detailed balance is satisfied (Appendix \ref{app:check_db}).  Equilibrium dynamical (i.e. two-time) quantities are averaged both over time and over 20 (high $T$) or 40 (low $T$) independent initial conditions from the equilibrium distribution.  Aging simulations are instead averaged over 50 independent initial conditions.

\section{Comparison of TRSP model with MD simulations}

\subsection{Dynamical heterogeneity}

To characterize the relaxation dynamics of the model we compute the overlap function, $Q{\left(t\right)}$, defined as

\begin{equation}
Q{\left(t\right)} = \frac{1}{N} \sum_i I_{i}{\left(t\right)},
\end{equation}
where $I_i{\left(t\right)}$ is an indicator variable that is $1$ if particle $i$ has not rearranged and $0$ if it has.   

The dynamical susceptibility is defined as

\begin{align}
    \chi_4{\left(t\right)}  = N \mathrm{Var} \,Q{\left(t\right)}.
\end{align}

We define the time at which $\chi_4$ is maximized as $\tau_\chi$ and its peak value to be $\chi_4^*$.

To test our model, we compare $Q$ and $\chi_4$ to observations from MD simulations. We adopt the picture that the dynamics consists of transitions between inherent states with vibrational motion superimposed on top ~\cite{monthus1996models,sastry1998}. Since our model only seeks to describe these inherent-state transitions, comparison to inherent-state $Q$ or $\chi_4$ is more natural than comparison to values computed using instantaneous positions. Using instantaneous positions, for example, reduces the peak value of $\chi_4$, which we interpret as effectively superimposing noise upon the real correlations in the dynamics.  

\begin{figure}[h]
    \includegraphics{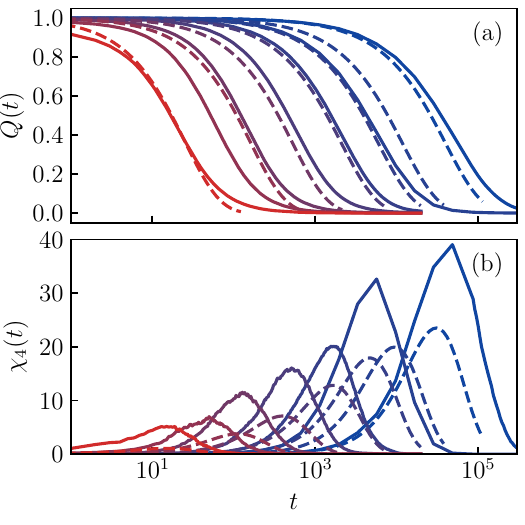}
    \caption{Comparison of TRSP model predictions (dashed) with MD simulations solid) at several different temperatures $T=0.42,0.45,0.47, 0.50, 0.55, 0.60, 0.70$, with the highest temperature in red. (a) Overlap function $Q(t)$.  (b) Magnitude of dynamical heterogeneity, $\chi_4(t)$.}  \label{fig:Qchi}
\end{figure}

\begin{figure}[h]
    \includegraphics{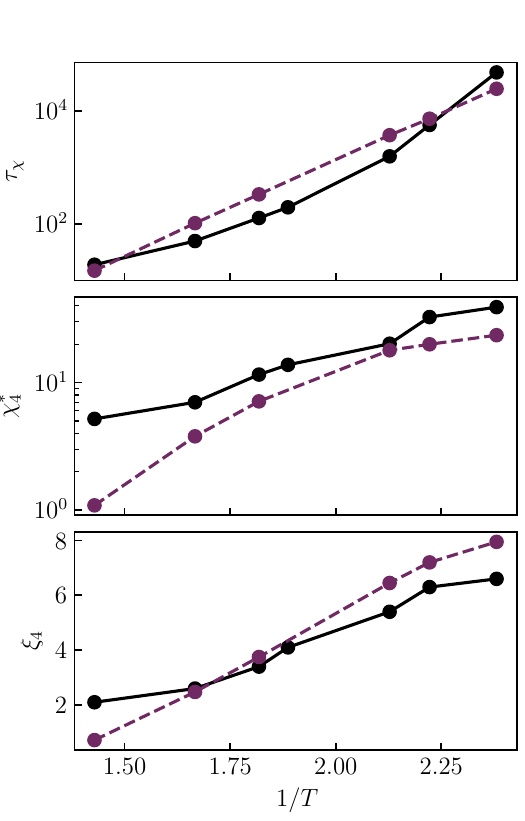}
    \caption{Quantitative comparison of dynamical heterogeneity predicted by the model (purple dashed) and measured from MD simulation (black solid). (a) The time $\tau_\chi$ at which $\chi_4(t)$ peaks. (b) Peak height $\chi_4^*$. (c) The dynamical length scale $\xi_4$. }\label{fig:chipeak}
\end{figure}

Figure \ref{fig:Qchi} shows $Q{\left(t\right)}$ and $\chi_4{\left(t\right)}$ for various temperatures $T$, showing qualitative agreement between MD simulations (A,C) and the model (B,D). Fig \ref{fig:chipeak}(A,B) show $\tau_\chi$ and $\chi_4^*$ for the MD simulations and the TRSP model as a function of $T$.   The agreement between the two is quantitatively reasonable. We note, however, that the TRSP model clearly underestimates the fragility (strength of non-Arrhenius behavior in $\tau_\chi$ vs. $T$) and the overall magnitude of $\chi_4^*{\left(T\right)}$. 

Note that with the exception of the highest temperature studied, $T=0.7$, the scaling of $\chi_4^*$ with $T$ is captured well by the model. The poor performance of the model near $T_0$ may be related to the breakdown of the assumption that the dynamics can be well-approximated by a sequence of discrete rearrangements. Alternatively, we note that the dynamical length scale at high $T$ approaches the lattice spacing. As a result, the observed deviation from our theory may arise from a breakdown of the lattice approximation or of the assumption that all rearrangements involve $m=4$ particles (see below).

For reference, note that a model of \textit{nonoverlapping} clusters of $m=4$ particles that rearrange together, each with identical rearrangement rates, would produce $\chi_4^* = m/4 = 1$~\cite{trap}.  The fact that the clusters in our model are allowed to overlap reduces $\chi_4^*$. relative to the trap prediction. Since all correlations of rearrangement rates vanish at $T_0$, the rough scale of $\chi_4^*$ near $T_0$ produced by the model matches our expectations. Thus, indeed, it is clear that the $\chi_4^*$ we see near $T_0$ in the MD data is too high to be explained by this model of discrete rearrangements. 

In supercooled liquids, the growth of $\chi_4^*$ with decreasing temperature is associated with growth of a dynamical correlation length $\xi_4$, which can become quite large at low temperatures. Since we have not included any growth of either the softness correlation length $\xi_S$ or the size $m$ of rearrangements in the model, it is not obvious that such a growing length scale is the origin of the growth of $\chi_4^*$ in our model.  To investigate this, we measure $\xi_4$ in both the TRSP model and MD simulations.

In each case, we measure $\xi_4$ at a single high temperature by fitting the dynamical structure factor $S_4(q, t)$, defined by

\begin{align}
\tilde{Q}(\mathbf{q},t) &= \frac{1}{N} \sum_i e^{i \mathbf{q} \cdot \mathbf{r}_i} I_i(t) \\
S_4(\mathbf{q}, t) &=  N \left( \langle \tilde{Q}(\mathbf{q},t) \tilde{Q}(-\mathbf{q},t) \rangle - \langle \tilde{Q}(\mathbf{q},t)\rangle \langle \tilde{Q}(-q,t)\rangle \right)
\end{align}
evaluated at $t=\tau_\chi$, to an Ornstein-Zernicke form~\cite{karmakar2010,flenner2010}, and then measure the growth of $\xi_4$ with cooling by performing a finite-size scaling collapse of $\chi_4^*$ measured in subsystems of varying size~\cite{tah2017block}. 

The result of this calculation is shown in Fig. \ref{fig:chipeak}(c). We see that, in fact, this model \textit{does} predict a dynamical length $\xi_4$ that grows with decreasing $T$, even though we have not included growth in the static length scale.  

How can the growing dynamical correlations $\xi_4(T)$ originate from a \textit{fixed} static length scale $\xi_S$? Intuitively, this is because a given difference in $S$ results in a bigger difference in rearrangement rates at lower $T$; thus, the same static correlations ``count for more'' at lower $T$. 

\begin{figure}[h]
    \includegraphics[]{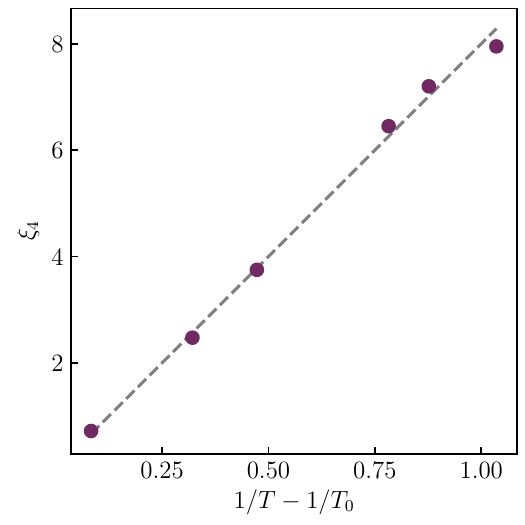}
    \caption{ $\xi_4^*(T)$ computed for the model as a function of $1/T-1/T_0$, showing a roughly linear dependence.   }
    \label{fig:xi_T0}
\end{figure}

More precisely, the contribution of $S$ to $\log{P_R}$ is proportional to $\gamma(T) \propto 1/T - 1/T_0$. Thus, in Fig. \ref{fig:xi_T0} we plot $\xi_4$ in our model as a function of $1/T - 1/T_0$. We find the relationship between the two to be nearly linear, supporting the picture that the dynamical length in this model comes, roughly speaking, from multiplying the static correlation length by this growing contribution of $S$ to the probability of rearrangement.

\subsection{Correlation vs.~causation in TRSP model} \label{sec:facdec}

We now turn to the origin of the growing dynamical correlations with decreasing $T$ within the TRSP model.

We can imagine three possible origins for the dynamics correlations in the model:

\begin{enumerate}
\item The constraint that particles rearrange in groups of $m$.
\item Spatial correlations in $S$ (correlation).
\item Facilitation, or causation ($\Delta S$ induced by rearrangements) 
\end{enumerate}

The first factor could be labeled as causation but really just accounts for the reality that a rearrangement involves more than one particle. It certainly makes a contribution---the single-site rearrangement version of the TRSP model produces a much smaller value of $\chi_4^*$ at all temperatures (see Appendix \ref{app:singlesite}). However, although this affects the magnitude, it does not affect the temperature-dependence significantly. We therefore focus on the second and third effects.

\begin{figure}[h]
    \includegraphics[]{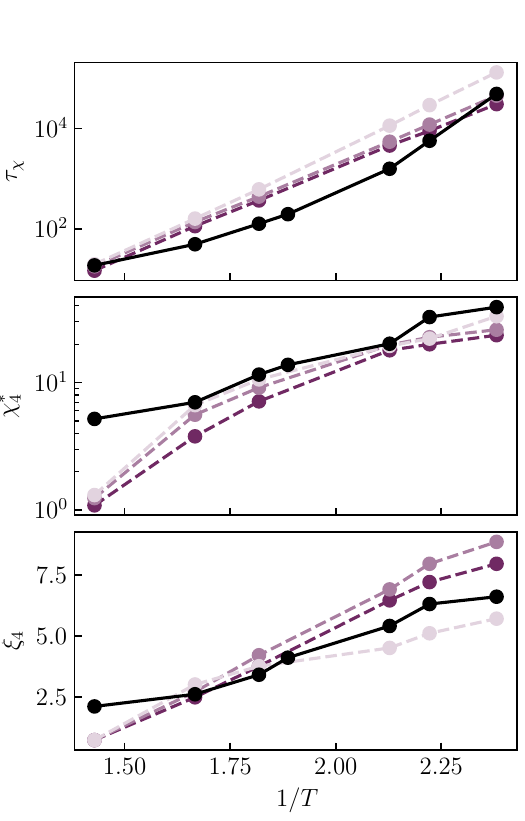}
    \caption{Effect of weakening facilitation on dynamical heterogeneity. Black dashed lines are MD data, and dark purple is model, as in Fig. \ref{fig:chipeak}. Medium and light purple correspond to reducing the variance of $\Delta S$ to $60\%$ or $20\%$ of its value ($c=0.6, 0.2$). (a) Weakening facilitation raises $\tau_\chi$. (b) Weakening facilitation increases $\chi_4^*$, at least at the current system size $L=20$. (c) Weakening facilitation to $c=0.6$ increases $\xi_4$ at the values of $T$ studied, but weakening facilitation to $c=0.2$ actually decreases $\xi_4$ at low $T$.}
    \label{fig:weak_facil}
\end{figure}

One virtue of the TRSP model is that the contributions of correlation and causation to dynamical heterogeneity are transparent. To investigate the relative roles of (2) and (3), we take advantage of our explicit conditions for guaranteeing time-reversal symmetry to generate alternate models. Using the detailed-balance conditions, we can construct a more general model in which the variance of $\Delta \mathbf{S}$ for all particles is reduced by a factor $c$, while holding the spatial correlations of $\mathbf{S}$ fixed. Thus, the contribution to $\chi_4$ from static spatial correlations in mobility is held fixed, but the effect of rearrangements on future mobility (facilitation) is weakened--correlation is held fixed by causation is reduced by a factor $c$. The result of this analysis is shown in Fig. \ref{fig:weak_facil}. Evidently, weakening facilitation in this model increases both $\tau_\chi$ and $\chi_4^*$. Thus, it appears that the growth of dynamical correlations in this model arises from the static softness correlations $\xi_S$, and facilitation (that is, changes in $\Delta S$ induced by rearrangements) only \textit{reduces} $\chi_4^*$. 

The fact that facilitation \textit{lowers} $\chi_4^*$ may seem surprising at first but is intuitively plausible.
  Recall that facilitation must satisfy detailed balance. Thus, despite what one might imagine from the word ``facilitation'', it is just as likely to make a neighboring particle \textit{less soft} as it is to make it \textit{softer}. One can imagine that the effect of facilitation on $\chi_4$ depends on temperature. At sufficiently high $T$, the neighbor of a soft particle is likely already soft due to the equilibrium spatial correlations. Indeed, it is likely soft enough to rearrange with high probability, even without facilitation. If faciltation lowers its softness, however, it will weaken the dynamical effects of the correlation present in the initial structure.

From this reasoning, it seems plausible that the sign of the effect may change at low enough $T$, when the average softness is lower and so that particles are unlikely to rearrange even if they are soft. In this case, lowering $S$ of nearby particles may effectively make no difference, while raising it may allow an avalanche to propagate.  Indeed, Figure \ref{fig:weak_facil}(c) shows that weakening facilitation actually lowers $\xi_4$ at low $T$, and in the finite-size collapse (Appendix \ref{app:collapse}, Figure \ref{fig:model_xi_collapse_2}), the extrapolation of $\xi_4^*$ to infinite system size actually decreases at $c=0.2$ for low $T$ (data not shown). Thus, at low $T$, it is possible that TRSP calculations for larger systems (which could be achieved by modifying the code to cut off the change in $\Delta S$ beyond a certain range) may show a crossover in the effect of facilitation on $\xi_4^*$.  Similarly, finite size effects may contribute to the lack of fragility observed, since the finite system size will cut off the growth of $\chi_4(t)$ due to growth of mobile domains when their size becomes comparable to the system size.

\subsection{Aging}

We next investigate the behavior of our model during aging. Although detailed balance is not satisfied out-of-equilibrium, this does not preclude using our model to study aging---indeed, the true dynamics do not satisfy detailed balance out of equilibrium, but crucially, out of equilibrium MD is still governed by the same dynamical equations as in equilibrium. Aging should still be described by dynamical equations which satisfy detailed balance once the steady state is reached.

\begin{figure}[h]
    \includegraphics[]{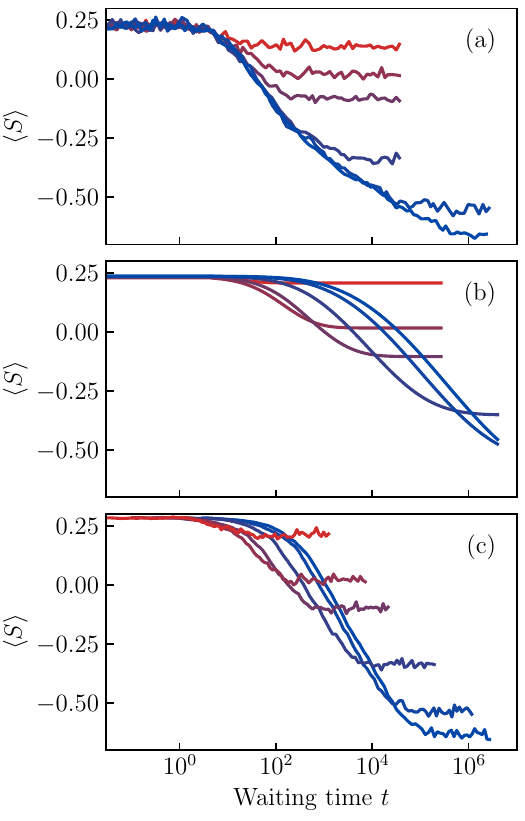}
    \caption{Aging dynamics of $S$. (a) MD data, showing collapse of $\langle S(t) \rangle$ at intermediate times. (b) A trap-like model of independent particles fails to reproduce the collapse of $\langle S(t) \rangle$. (c) Our model also fails to reproduce the collapse but does better than the trap model. }
    \label{fig:aging}
\end{figure}

In the MD simulations, the aging dynamics of $\langle S \rangle$  as a function of waiting time $t$ show a remarkable phenomenon: for samples quenched from the same high temperature $T_\mathrm{I}$ to \textit{different} final temperatures $T_\mathrm{F}$, $\langle S(t)\rangle$, is the same at intermediate times (Fig. \ref{fig:aging}(a) and~\cite{schoenholz2017relationship}).

This collapse cannot be reproduced by a model without interactions between particles~\cite{trap}. Intuitively, this is because, if $S(t)$ is the same at some time $t$ for different quench temperatures $T_\mathrm{F}$, then the colder temperature will have slower aging dynamics and thus the $\langle S(t)\rangle$ curves must separate as $t$ increases.

We modified the trap-like, independent particle model of~\cite{trap} to have a $T$-dependent density of traps, in order to produce the correct distribution of $S$ at every $T$. We then simulate aging in this model, showing $\langle S(t)\rangle$ in (\ref{fig:aging}(b)). We find that it still fails to produce the collapse of $\langle S(t)\rangle$ during aging which is observed in MD data. Thus, the trap model cannot reproduce the observed MD behavior.

The same calculations of $\langle S \rangle$ can be carried out on the TRSP model during aging (Fig. \ref{fig:aging}(c)). We find that including interactions between sites as in the TRSP model still fails to reproduce the collapse of $\langle S(t)\rangle$ at intermediate times seen in the MD data (Fig. \ref{fig:aging}(a)). However, the different curves are much closer together than in the trap model, suggesting that perhaps the interactions between sites are not sufficiently strong in the TRSP model.

\section{Discussion}

\subsection{Underestimation of fragility and \texorpdfstring{$\chi_4$}{dynamical heterogeneity}}

We now consider possible explanations for why the TRSP model underestimates fragility and $\chi_4$. Past work explaining $Q(t)$ empirically using softness~\cite{schoenholz2016structural} included two corrections that we have neglected here. First, they measured the fraction of rearrangements that involve a displacement greater than the threshold used to compute $Q$ at each $T$, and corrected the prediction by this factor. Secondly, they measured the fraction of rearrangements that reverse themselves at each $T$, and corrected the prediction by this factor. Both of these corrections have the effect of increasing $\tau$ at low $T$, while having little effect at high $T$. Thus, they tend to increase fragility. We could have included these effects but chose not to, in order to present the model in its simplest form.

Another likely reason for the underestimation of fragility and $\chi_4$ is the shortcoming of $S$ as a structural descriptor.  More complex machine learning methods produce structural descriptors that correlate more strongly with long-time dynamics than the linear method used to define $S$~\cite{jung2022predicting,bapst2020unveiling,boattini2021averaging,alkemade2023improving}. If such methods are optimized to predict short-time rather than long-time dynamics, they would presumably result in stronger dependencies of $\Delta E$ and $\Sigma$ on the descriptors, increasing both the degree of heterogeneity and fragility produced by the model.  Indeed, the range of $\Delta E$ and $\Sigma$ produced by a structural descriptor is likely to be a good indicator of the quality of that descriptor. For example, the structural variable constructed using a neural network in~\cite{jung2022predicting} seems to correlate extremely well with dynamical heterogeneity and could be a better short-time descriptor than softness. (Note, however, that the $\chi_4(t)$ reconstructed in that work using their structural variable is not the same as our $\chi_4(t)$, as it neglects certain kinds of dynamical fluctuations.) 

These considerations point to the importance of testing the Arrhenius dependence for $P(R)$ for structural descriptors besides softness. A TRSP-type model could be constructed using any scalar descriptor of structure. All that is required is that the dependence of the probability of rearrangement is Arrhenius for particles with a given value of the structural descriptor, and the distribution of values of the descriptor is reasonably Gaussian. This is why we regard the TRSP approach as a general framework for constructing time-reversal-invariant structuro-plasticity models.

A third contributing factor is that we have only included the effects of rearrangements of large $A$ particles in our binary $AB$ system on other $A$ particles. The $A$ particles belong in the majority, but rearrangements of small $B$ particles are actually more numerous, especially as $T$ decreases, and can facilitate rearrangements of both $A$ and $B$ particles. 

Finally, recall that there is evidence that the range of spatial correlations of $S$ increases with with decreasing $T$. Such an increase would presumably also lead to greater fragility and larger values of $\chi_4$.

\subsection{The role of elasticity}

As noted in the introduction, we have neglected elastic (far-field) facilitation. On timescales much shorter than $\tau_\alpha$ one expects the supercooled liquid to behave as a solid (as indeed glasses do when observed out of equilibrium). Furthermore the inherent state is, by construction, a forced-balanced state and therefore perturbations around it should be described on long length scales by continuum elasticity.  Recent work has demonstrated clear anisotropies in the correlated dynamics of supercooled liquids below the mode-coupling crossover temperature $T_{\mathrm{MCT}}$, which provide evidence that at least some facilitation arises from elasticity at sufficiently low temperatures~\cite{rahulMCT}.  Thus at low $T$ we expect that this model should pick up terms which make it resemble in spirit a thermal elastoplastic model with a softness kernel in addition to an elastic kernel, in a manner similar to ~\cite{Zhang2021,Zhang2022,Xiao2023}. 

It is not entirely obvious whether or not far-field elastic facilitation should increase or decrease $\chi_4^*$. On one hand, on short timescales, elastic facilitation, by pushing neighbors over their thresholds simultaneously, effectively acts like our ``multi-particle rearrangements'' ($m$), which should increase $\chi_4^*$. On the other hand, elastoplasticity also produces a long-time facilitation effect, which in our current model tends to decrease $\chi_4^*$. Thus the net effect of elastoplasticity on $\chi_4$ is unclear. On the empirical side, recent very-low $T$ simulations of the Kob-Andersen Lennard-Jones system appear to display a strong-to-fragile crossover with a slowing growth of $\chi_4^*$ at low $T$, which could be consistent with elastoplastic-facilitation-weakened heterogeneity~\cite{das2022}.

Recently, a simple thermal EPM has been studied as an explanation of dynamical heterogeneity~\cite{Misaki2022}. In this model, which does not include near-field structural facilitation, elastic facilitation alone is found to produce a growth of $\chi_4^*$ with decreasing temperature.  However, there are two factors which make it unclear how this result will generalize to more realistic models. First, because this model reduces to the trivial ``trap" limit of uncorrelated sites when elasticity is removed, $\chi_4^*$ cannot, by construction, decrease when elasticity is incorporated\footnote{We do not have a rigorous proof of this fact, but it seems very likely that the case of uncorrelated, identical barriers provides a lower bound on the degree of dynamical correlation.}. Second, the thermal EPM~\cite{Misaki2022} does not respect detailed balance. As we have argued in section \ref{sec:facdec}, our observation that facilitation decreases $\chi_4^*$ seems to be intimately connected to detailed balance, and indeed it is intuitively obvious that if $\Delta \mathbf{S}\left(\mathbf{r}\right)$ for some particular $\mathbf{r}$ were strictly positive then it would increase the degree of dynamical correlation.  Put differently, while one may interpret Ref.~\cite{Misaki2022} as showing that elastoplastic facilitation increases $\chi_4^*$, it is doing so with the spatial correlations of the barriers not held fixed, but instead being allowed to increase by the introduction of elasticity. Note, however, that it has been argued that the origin of the dynamic length scale in these models is a ``nucleation and growth'' process rather than any spatial correlations of barriers ~\cite{tahaei2023scaling}. It would be extremely interesting to investigate these questions in a thermal EPM that respects time-reversal symmetry. 

Another recent model, composed of discrete excitations interacting elastically, very naturally respects time reversal symmetry, and may suggest an alternate route to adding elastic interactions to our model, as well as another model in which to study the effects of facilitation on $\chi_4$~\cite{haysim2023}.

As noted in~\cite{Misaki2022}, however, it remains true that dynamic heterogeneity with similar phenomenology is observed in systems outside of thermal equilibrium, e.g. granular systems~\cite{keys2007,abate2007topological}.  Thus, although the details of how e.g. facilitation contributes to dynamical heterogeneity may depend on time-reversal symmetry as we have argued, the basic phenomenolgy must remain. It is unclear what principles may be used to constrain our models of the dynamics of $S$ when we depart from thermal equilibrium.

\section{Acknowledgements}
We thank Rahul Chacko, Indrajit Tah, Richard Stephens, Misaki Ozawa, Muhammad Hasyim, and Matthieu Wyart for helpful discussions. This work was supported by the Simons Foundation via the ``Cracking the glass problem'' collaboration (\#454945, SAR and AJL), and Simons Investigator awards to AJL (\#327939) and to Ilya Nemenman (SAR). AJL thanks CCB at the Flatiron Institute, as well as the Isaac Newton Institute for Mathematical Sciences under the program ``New Statistical Physics in Living Matter" (EPSRC grant EP/R014601/1), for support and hospitality while a portion of this research was carried out.

\appendix
 
\section{Structure functions used to construct softness} \label{app:sf}

We use the same structure functions as in~\cite{trap}. These are the same as those used in~\cite{cubuk2015identifying,schoenholz2016structural,schoenholz2017relationship}, but with more radial structure functions. We use the radial structure functions

\begin{equation}
G{\left(i;  \mu, \sigma\right)} = \sum_{ j | r_{ij} < \sigma_{\mathrm{max}}} e^{-\left(r_{ij} - \mu\right))^2 /\sigma^2 }.
\end{equation}

We take $\sigma = 0.05 \sigma_{AA}$, $\mu = 0.05 \sigma_{AA}, 0.1 \sigma_{AA}, \dots, 5 \sigma_{AA}$, and $\sigma_{\mathrm{max}} = 5 \sigma_{AA}$

We also use the angular structure functions

\begin{align}
&\Psi{\left(r; \xi, \lambda, \zeta \right)} =\nonumber  \\ &\sum_{ j,k | r_{ij}, r_{jk} < \sigma_{\mathrm{max}}} e^{-\left(r_{ij}^2 + r_{jk}^2 + r_{ik}^2\right)/\xi^2 } \left( 1 + \lambda  \cos{\theta_{ijk}}\right)^{\zeta},
\end{align}

with $\xi$, $\lambda$, and $\zeta$ taking $66$ combinations of values, as in~\cite{cubuk2015identifying}.

\section{Details of detailed-balance condition derivation} \label{app:fulldb}

As stated in the main text we consider a lattice model, where the vector $\mathbf{S}$ contains the softness of each site. We assume that at equilibrium $\mathbf{S}$ is drawn from a multivariate normal distribution, with a covariance matrix $\cov$, i.e. 

\begin{equation}
P{\left(\mathbf{S}\right)} \propto e^{-\left(\mathbf{S} - \mu \right) \cov^{-1} \left(\mathbf{S} - \mu \right)^T }.
\end{equation}

The probability of a particle rearranging depends on its softness through

\begin{equation}
P{\left(R|S\right)} = A{\left(T\right)} e^{\gamma{\left(T\right)} S - \delta{\left(T\right)} S^2},
\end{equation}

where the temperature dependence of the coefficients is Arrhenius. The correction term $\delta S^2$ is small. 

To determine how detailed balance constrains the distribution of $\Delta S$ for a rearrangement, we write down the distribution of $\mathbf{S}$ given that a particle at the origin is rearranging:

\begin{equation}
    P{\left(\mathbf{S} | R\right)} \propto P{\left(R | S{\left(0\right)} \right)} P{\left(\mathbf{S}\right)},
\end{equation}

a normal distribution whose covariance we denote $\covR$.  Taking the logarithm of both sides of this equation (note that the quantities inside the exponentials are just real numbers) yields (with $C$ an arbitrary constant, and $\zv$ the vector with a $1$ at site $0$ and zeros elsewhere):

\begin{widetext}
\begin{align}
-\left(\mathbf{S} - \muR  \right)^T \covR^{-1} \left(\mathbf{S} - \muR \right) + C&= 2 \gamma S(0) - 2 \delta S(0)^2 -\left(\mathbf{S} - \mu  \right)^T \cov^{-1} \left(\mathbf{S} - \mu \right) \label{eq:cov_derivation_1}\\
&=2 \gamma S(0) - 2 \delta \mathbf{S}^T \zv \zv^T \mathbf{S}  -\left(\mathbf{S} - \mu  \right)^T \cov^{-1} \left(\mathbf{S} - \mu \right) \\
&= 2\left(\gamma - 2 \delta\mu\right) S(0) +2 \delta \mu^2 -\left(\mathbf{S} - \mu  \right)^T  \left(\cov^{-1} + 2 \delta \zv \zv^T \right)  \left(\mathbf{S} - \mu \right).
\end{align}

Now introduce the covariance matrix of rearranging particles, $\covR = \left(\cov^{-1} +2 \delta \zv \zv^T  \right)^{-1}$, and write

\begin{align}
-\left(\mathbf{S} - \muR  \right)^T \covR^{-1} \left(\mathbf{S} - \muR \right) + C&=  2\left(\gamma - 2 \delta\mu\right) \mathbf{S}^T \hat{0} \covR \covR^{-1} +2 \delta \mu^2 -\left(\mathbf{S} - \mu  \right)^T \covR^{-1} \left(\mathbf{S} - \mu \right) \\
&= C + \left(\mathbf{S} - \mu  - \left(\gamma - 2\delta \mu\right) \covRij{0}{i} \right)^T   \covR^{-1} \left(\mathbf{S} - \mu  - \left(\gamma - 2\delta \mu\right) \covRij{0}{i} \right). \label{eq:cov_derivation_2}
\end{align}
\end{widetext}

We then immediately read off the mean softness at distance $r$ from a rearranging particle, and use the Sherman-Morrison formula to compute $\covR = \left(\cov^{-1} +2 \delta \zv \zv^T  \right)^{-1}$. 
This gives us the mean softness at a distance $r$ from a rearranging particle, and $\covR$,  as

\begin{align}
    \langle \mathbf{S}_i\rangle \equiv \muR &= \mu + \left( \gamma -2 \delta \mu\right) \covR_{0,i} \\
    \covRij{i}{j} &= \covij{i}{j} - \frac{2 \delta\covij{0}{i} \covij{0}{j}}{1 + 2\delta \cov_{00}}.
\end{align}

Now  we assume that $\Delta \mathbf{S}$ is Gaussian with mean

\begin{equation}
\langle \Delta \mathbf{S} \rangle = \mathbf{H}\left( \mathbf{A}- \mathbf{S} \right), \label{eq:applinear}
\end{equation}

and covariance $\sigds$.

In the main text we asserted that this assumption gives a distribution of $\left(\mathbf{S}, \mathbf{S}' = \mathbf{S} + \Delta \mathbf{S} \right)$ which is Gaussian. To see this, write
\begin{widetext}
\begin{align}
P{\left(\mathbf{S}, \mathbf{S}' | R\right)} &= P{\left(\mathbf{S} | R \right)} P{\left(\Delta \mathbf{S} | R, \mathbf{S}\right)} \\
 &= \propto \exp(-\frac{1}{2}\left(\mathbf{S} - \muR  \right)^T \covR^{-1} \left(\mathbf{S} - \muR \right))  \exp(-\frac{1}{2} \left(\mathbf{S}' - \mathbf{S} - \mathbf{H} \left( \mathbf{A}- \mathbf{S} \right)\right)^T \sigds^{-1} \left(\mathbf{S}' - \mathbf{S} - \mathbf{H} \left( \mathbf{A}- \mathbf{S} \right)\right) ) \\ 
 &= \exp(-\frac{1}{2} \left[\left(\mathbf{S} - \muR  \right)^T \covR^{-1} \left(\mathbf{S} - \muR \right)  +  \left(\mathbf{S}' - \mathbf{S} - \mathbf{H} \left( \mathbf{A}- \mathbf{S} \right)\right)^T \sigds^{-1} \left(\mathbf{S}' - \mathbf{S} - \mathbf{H} \left( \mathbf{A}- \mathbf{S} \right)\right) \right])
\end{align}
\end{widetext}

What appears inside the exponential is a quadratic form in the vector $\left(\mathbf{S}, \mathbf{S}'\right)$ and thus the joint distribution is Gaussian. As sketched in the main text we will make use of this fact to facilitate our derivation of the detailed-balance conditions.  We use the fact that $\mathbf{S}$ and $\mathbf{S}'$ have a joint Gaussian distribution in the case of a linear correlation between $\Delta \mathbf{S}$ and $\mathbf{S}$, and describe this joint distribution by the block covariance matrix of the random variable $\tilde{\mathbf{S}}= \left( \mathbf{S}, \mathbf{S}'\right)$:

\begin{align}
    \tilde{\mathbf{K}} &= \left( \begin{array}{cc} \covR & \tcor \\ \tcor & \covR \end{array}\right).
\end{align}

The fact that the rearrangement obeys time-reversal symmetry is expressed in the equality of the diagonal and off-diagonal blocks. Since a linear transformation $Y = A X$ of a Gaussian random variable $X$ with covariance $\Sigma_X$  and mean $\langle X \rangle$ yields $\Sigma_Y  = A \Sigma_X A^T$ and $\langle Y \rangle = A \langle X \rangle$, we construct the covariance and mean of $\mathbf{y} = \left(\mathbf{S}, \Delta \mathbf{S}\right)$

\begin{align}
    \langle \mathbf{y} \rangle &= \left(\muR, \mathbf{0}\right) \\
    \Sigma_y &= \left(\begin{array}{cc}
    \covR & \tcor - \covR \\ \tcor  - \covR & 2 \left(\covR - \tcor\right)
    \end{array} \right).
\end{align}

Finally, we condition on $\mathbf{S}$ to construct $P{\left(\Delta \mathbf{S} | \mathbf{S}\right)}$.  For $\left(\mathbf{x}_1, \mathbf{x}_2\right)$ with a joint Gaussian distribution, it is known that $P{\left(\mathbf{y}|\mathbf{x}\right)}$ is Gaussian with covariance $\mathbf{K}_{22} - \mathbf{K}_{21} \mathbf{K}_{11}^{-1} \mathbf{K}_{12}$ and mean $\boldsymbol{\mu}_2 + \mathbf{K}_{21} \mathbf{K}_{11}^{-1} \left(\mathbf{x} - \boldsymbol{\mu}_x\right)$. Thus, $P{\left(\Delta \mathbf{S} | \mathbf{S}\right)}$ is Gaussian with

\begin{align}
\left\langle \Delta \mathbf{S}\right\rangle &= \left(\tcor - \covR\right)  \covR^{-1}\left(\mathbf{S} - \muR\right)\\
&= \left( 1\!\!1  - \tcor\covR^{-1}\right) \left( \muR - \mathbf{S} \right) \\
&\equiv \mathbf{H}\left(\muR - \mathbf{S}\right) \\
\sigds &= 2 \left(\covR - \tcor\right) -  \left(\tcor - \covR \right) \covR^{-1} \left(\tcor - \covR \right) \\
&=  2 \mathbf{H}\covR  - \mathbf{H} \covR \covR^{-1} \mathbf{H}\covR \\
&= \left(2 - \mathbf{H}\right) \mathbf{H}\covR.
\end{align}

which  this yields the conditions

\begin{align}
    \mathbf{A} &= \muR \\
    \sigds &= \left(2 - \mathbf{H} \right) \mathbf{H}\covR,  \label{eq:appcovar}
\end{align}

as claimed in the main text. As stated in the main text, $\sigds$ satisfying this equation is given by Equation \ref{eq:sqrt}.

Working directly with detailed balance instead, after a substantial amount of algebra, yields the condition

\begin{equation}
\sigds^{-1} - \covR^{-1} = \left(1\!\!1 - \mathbf{H}^T \right) \sigds^{-1} \left(1\!\!1- \mathbf{H} \right). \label{eq:db}
\end{equation}

$\mathbf{H}$ given by Equation \ref{eq:sqrt} (i.e. the solution to equation \ref{eq:appcovar}) also solves equation \ref{eq:db}.  This follows from the fact  that this solution is of the form $\mathbf{H} = f{\left(\sigds \covR^{-1}\right)}$ for an analytic function $f$. By expanding $f$ in series and using the symmetry of $\sigds$ and $\covR$ we may show that $ \sigds^{-1} f{\left(\sigds \covR^{-1}\right)}$ is a symmetric matrix for any analytic $f$; with this fact it is straightforward to manipulate equation \ref{eq:db} to be the same as equation \ref{eq:appcovar}.

Also note that, to obtain equations \ref{eq:covmult}--\ref{eq:meanmult} in the main text, one may simply replace $\zv$ with a vector that has value $1/m$ on the $m$ rearranging sites in equations \ref{eq:cov_derivation_1}--\ref{eq:cov_derivation_2}.

\section{Choice of multi-site rearrangement rule} \label{app:kcorr}

To generalize the observed $P{\left(R|S\right)}$  and $P{\left(\Delta \mathbf{S}\right)}$into a rule for multi-body rearrangements we must satisfy two criteria:

\begin{enumerate}
    \item The size of elementary rearrangements should be independent of $T$, or perhaps weakly dependent on $T$. In particular elementary rearrangements should not involve $100$ times as many particles just because the relaxation time is $100$ times shorter. \label{it:size}
    \item Time-reversal symmetry should remain satisfied for $P{\left(\Delta \mathbf{S}\right)}$ \label{it:trev}
    \item Ideally, the different particles participating in the rearrangement should be treated in the exact same way. At the very least, the identity of all rearranging particles should affect $P{\left(\Delta \mathbf{S}\right)}$, to preserve dynamical correlations. \label{it:perm}
\end{enumerate}

The most obvious way to satisfy condition \ref{it:perm}  is to simply make the probability of a multi-particle rearrangement involving particles $i_1, i_2, \dots, i_m$ equal to $P{\left(R | S_{i_1}) \right)} P{\left(R | S_{i_2}) \right)} \cdots P{\left(R | S_{i_m}) \right)}$ This, however, runs afoul of condition \ref{it:size}: it will produce single-particle rearrangements at low temperatures and many-particle rearrangements at high temperatures.  

This naturally leads us to consider rules where the ``secondary'' rearrangements $S$ is not involved in the probability of rearrangement---i.e. the rearrangement of a ``soft'' particle can easily force a ``hard'' neighbor to go along with it.

The simplest way to get a rule which satisfies condition \ref{it:trev} and \ref{it:size} is to simply determine a ``primary'' rearranging particle according to $P{\left(R|S\right)}$, determine simultaneous ``secondary'' rearranging particles according to some other rule, and only use the primary rearranging particle to determine $P{\left(\Delta \mathbf{S}\right)}$.  This is the approach of reference~\cite{Xiao2023}.  This is, however, a strong violation of condition \ref{it:perm}.  

One other idea, then, is to trigger secondary rearrangements with a softness-independent probability $f{\left(r\right)}$ as in reference~\cite{Xiao2023}, but then allow each of these secondary rearrangements to produce some $\Delta \mathbf{S}$.  This still does not give full permutation symmetry between the different rearranging particles (the ``primary rearrangement'') is singled out), but would still preserve some dynamical correlations caused by which rearranging neighbour rearranges.  It is difficult, however, to ensure that such a scheme truly satisfies time-reversal symmetry.  In such a scheme, the simultaneous rearrangement of particles $A$ and $B$ can proceed by one of two ``pathways'': one dependent on $S_A$ and one dependent on $S_B$. If we assume that these two different processes produce the same distribution of $\Delta \mathbf{S}$ then the equations of detailed balance become analytically intractable. On the other hand, if we assume the two processes have different distributions of $\Delta \mathbf{S}$ we find that full symmetry between the two single-particle rearrangements still cannot be restored: in order to satisfy detailed balance the distribution of $\Delta \mathbf{S}$ for the two rearrangements must be different. 

The simplest rule which satisfies all the conditions is to use the average $S$ of the rearranging sites in order to determine the probability of  a rearrangement. The disadvantage of this rule, however, is that it offers no insight into what sets the size of elementary rearrangements: we must put the distribution of rearrangement sizes $m$ in by hand, and in our case we choose to fix the size.

We then must fix the distribution of  $\Delta \mathbf{S}$  in order to satisfy condition \ref{it:trev}. Using the average $S$ still permits this as it keeps $P{\left(R|S\right)}$ Gaussian; a rearrangement probability which is exponential in any second-order polynomial of $S_1, S_2, \dots$ will work. One might hope that there is a straightforward way to ``build''  the distribution of $\Delta \mathbf{S}$ by the successive action of two single-particle rearrangements.   If there was a method which allowed use to think of the two rearrangements as ``simultaneous'' (e.g. if $\Delta \mathbf{S}$ had a distribution which was independent of the order of rearrangements) then we could proceed as such.  One finds however, that the distribution of $\Delta \mathbf{S}$ in such a case depends on the \textit{order} of the rearrangements. Thus, if we wish to satisfy time-reversal symmetry we require the rearrangement ``particle $A$ rearranges, then particle $B$ rearranges'' to proceed by two independent pathways, one determined by $S_A$ and one by $S_B$, with no reference at all to the ordering of the rearrangements.  Thus, the most straightforward approach is to construct $\sigma_\Delta$ with no reference to hypothetical single-particle rearrangements, e.g. by assuming it is a function of distance from the center of mass of the rearrangement, or from the set of all rearranging particles as we have done.

\section{Renormalization of \texorpdfstring{$P{\left(R|S\right)}$}{rearrangement probability} for multi-site rearrangements} \label{app:renorm}

As in the main text, assume that each group of $m$ neighbouring sites rearranges with a probability \begin{equation}
P{\left(R_{\left\{i_1,\dots i_m\right\}}|S, T\right)} = P_0{\left(R|\langle S\rangle_m, T\right)},
\end{equation}

for some ``bare'' rate $P_0$. The apparent rearrangement rate for each particle is thus

\begin{align}
\frac{\bar{P}{\left(R|S_{i_1}\right)}}{z_m} =\int \dd^{m-1}\mathbf{S}_{\left\{i_2, \dots, i_m\right\}}  P{\left(\mathbf{S}_{\left\{i_2, \dots, i_m\right\} }| S_{i_1} \right)} P_0{\left(R| \langle S\rangle_m \right)},
\end{align}

where $z_m$ is the number of clusters of size $m$ we consider for rearrangement that contain a given site (e.g. $z_2=6$ and $z_4=12$ on a cubic lattice).

We note that the conditional distribution $P{\left(\mathbf{S}_{\left\{i_2, \dots, i_m\right\} }| S_{i_1} \right)}$ is Gaussian. Its  covariance is $\cov_{i_j, i_k} - \cov_{i_j, i_1}\cov_{i_k, i_1} / \sigma^2 $, and its mean is $\mu_{i_j} = \langle S \rangle + \cov_{i_1, i_j} \left(S_i - \langle S \rangle \right)$. 

Doing the Gaussian integral yields the observed $P{\left(R|S\right)}$ in terms of the bare parameters. The result is of the form

\begin{align}
P{\left(R|S\right)} &= A e^{\gamma S- \delta S^2},
\end{align}

with the ``renormalized'' parameters $A, \gamma, \delta$ determined by the ``bare'' parameters $A_0, \gamma_0, \delta_0$.  The set of equations relating them may be solved to obtain $A_0, \gamma_0$, and $\delta_0$. This produces the equations in the main text.

\section{Subtracting off the background from measurements of \texorpdfstring{$\sigma$ and $\eta$}{softness changes}} \label{app:subtract}

As mentioned in the main text, we interpret the values of $\sigma^2_{\Delta}{\left(r\right)}$ and $\eta{\left(r\right)}$ as $r\to \infty$ in the MD data as a background value caused by undetected rearrangements. Here we explain in detail how the background events are subtracted off.

Intuitively, if we imagine that the observed $\Delta S$ is a sum of two processes, $\Delta S_1$ and $\Delta S_2$, we might suppose that we can take advantage of the linearity of the mean and variance, to write $\eta = \eta_1 + \eta_2$ and $\sigma^2_{R} = \sigma^2_{R,1} + \sigma^2_{R,2}$.  Thus, we could obtain the parameters for the rearrangement by subtracting off the background. This produces results which look reasonable at long distances and low temperatures, but it cannot be consistent if $\sigma^2_{\Delta}$ is too large (it is unphysical to have $\eta > 1, \sigma_\Delta > \sigma$.)

In reality, $\Delta S_1$ and $\Delta S_2$ cannot be assumed to be uncorrelated, since the mean $\Delta S_2$ is correlated with $S + \Delta S_1$.  If we assume a gaussian distribution of $\Delta S$ and do the integral over the possible values of $\Delta S_1$, then substitute in the approximate (uncorrelated) detailed balance condition \ref{eq:nocorr}, we arrive at the corrected sum formulae 

\begin{align}
\eta &= \eta_1 + \eta_2 - \eta_1 \eta_2  \\
\sigma_\Delta^2 &= \sigma_{R,1}^2 + \sigma_{R,2}^2 - \frac{\sigma_{R,1}^2 \sigma_{R,2}^2}{\sigma^2}.
\end{align}

These formulae are used to subtract off the background in Fig. \ref{fig:dS}.

In the above, we have ignored the effect of spatial correlations in $S$ and the covariance structure of $\Delta S$, under the assumption that this is sufficient for measuring the approximate magnitude of $\Delta S{\left(r\right)}$. Note that, as discussed in section \ref{app:kcorr}, if we wish to go beyond this approximation the idea of ``adding together'' the effect of multiple rearrangements becomes poorly defined, with the result depending on the ordering of the rearrangements.

\section{Tests of model implementation} \label{app:checks}

\subsection{Direct test of detailed balance} \label{app:check_db}

In equilibrium the model is constructed to satisfy detailed balance, i.e. $\psr{\mathbf{S}} \tp{\mathbf{S}}{\mathbf{S}'} =  \psr{\mathbf{S}'} \tp{\mathbf{S}'}{\mathbf{S}}$ for any $\mathbf{S}, \mathbf{S}' \in \mathbb{R}^N$.

Detailed balance is preserved by any coarsening of the state space, i.e. a mapping from $\mathbf{S}, \mathbf{S}' \in \mathbb{R}^N$ to $A, B$ in some discrete space. Thus, we may directly test detailed balance in our model simulations by letting $A$ represent the binned value of $S$ for a rearranging particle, and then checking that $\psr{A} \tp{A}{B} =  \psr{B} \tp{B}{A}$. 

\begin{figure}[h!]
    \centering
    \includegraphics[]{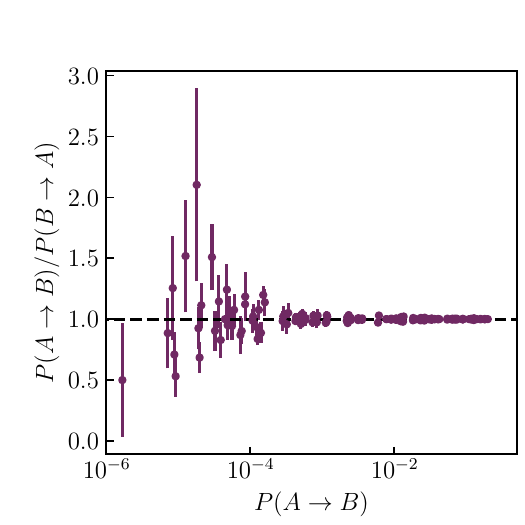}
    \caption{Check of detailed balance at $T=0.470$, for transitions between states defined by binning $S$ of a rearranging particle. Error bars are one standard deviation over bootstrap resamplings of independent simulation seeds. (The $x$-axis value also has an uncertainty, but its precise value doesn't matter.)}
    \label{fig:db_check}
\end{figure}

Figure \ref{fig:db_check} shows the results of this test, plotting $\psr{A} \tp{A}{B} / \psr{B} \tp{B}{A}$ against $\psr{A} \tp{A}{B}$ for the numerical simulation of the model at $T=0.470$. For this coarsening of the states, detailed balance is satisfied within error bars. 

\subsection{Mean and correlation functions of \texorpdfstring{$S$}{softness}} \label{app:check_mc}

The model is constructed to reproduce a known $\langle S \rangle$ and correlation function $\cf{\mathbf{r}}{\mathbf{r}'}$; we also verify this in our numerical implementation of the model.

Some care is required to precisely compute any such average over the distribution of $\mathbf{S}$. The samples produced by the Gillespie algorithm are of rearrangement events, and are thus biased toward their values for rearranging particles by fluctuations in $\langle S \rangle$ of order $\xi_S^3/ N$. To do proper time averaging, the samples obtained must be weighted by the time until the next rearrangement event. 

\begin{figure}[h!]
    \centering
    \includegraphics[]{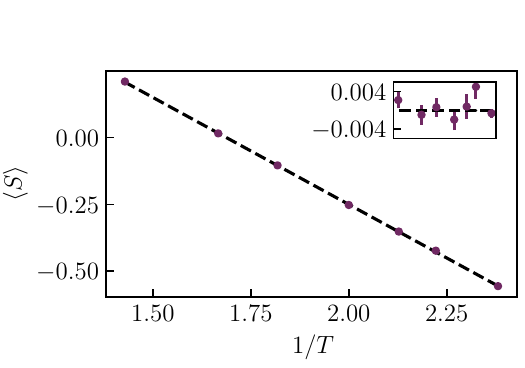}
    \caption{$\langle S\rangle$ as a function of $T$ for our model simulations (purple circles) matches the target (black dashed line). Error bars are one standard deviation over bootstrap resamplings of independent simulation seeds. Inset shows deviations between simulation and target, showing that deviations appears to be statistical noise.}
    \label{fig:mean_check}
\end{figure}

\begin{figure}[h!]
    \centering
    \includegraphics[]{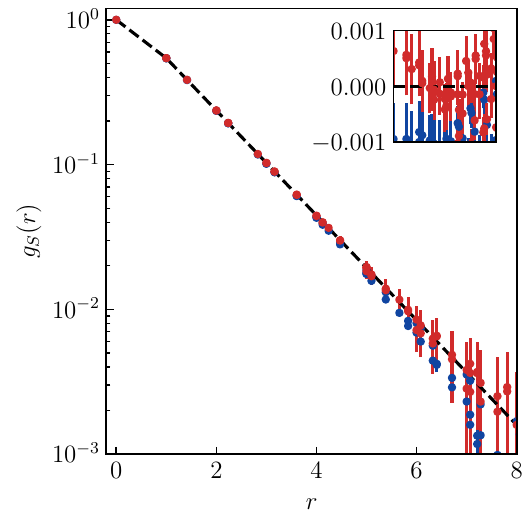}
    \caption{The correlation function of $S$, $C_S{\left(r\right)} = \left\langle \left(S{\left(r\right)} - \langle S \rangle \right)  \left(S{\left(0\right)} - \langle S \rangle \right) \right\rangle$ in our model simulations (blue, red circles) matches the target (black dashed line). Error bars are one standard deviation over bootstrap resamplings of independent simulation seeds.  Data are shown for our highest and lowest $T$, and each point represents a specific vector $\mathbf{r}$,  restricted to the plane $z=0$ for clarity. Inset shows deviations between simulation and target, showing that deviations appear to be statistical noise. (In a finite sample, fluctuations in a measured correlation function at different $r$ are anticorrelated.) }
    \label{fig:corr_check}
\end{figure}

Figure \ref{fig:mean_check} shows the simulated $\langle S \rangle$ as a function of the theoretical $\langle S \rangle$ for the used parameters, showing perfect agreement. Figure \ref{fig:corr_check} shows the correlation function $\cft{\left| \mathbf{r}-\mathbf{r}'\right|}$ at $T=0.420, 0.700$ alongside the target function, again showing perfect agreement. The same plot for other simulated temperatures (not shown) shows the same result, and the small deviations from the target correlation function do not show any systematic trend with temperature $T$.

\section{A model with single-site rearrangements strongly underestimates \texorpdfstring{$\chi_4^*$ }{dynamical heterogeneity}} \label{app:singlesite}

\begin{figure}[h!]
    \includegraphics[]{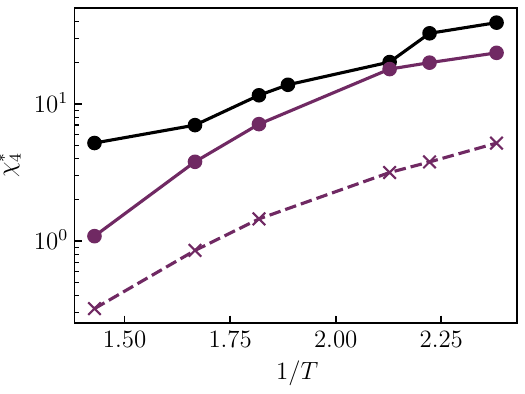}
    \caption{Model with single-site rearrangements (dashed line) substantially underestimates dynamical heterogeneity. Note that, in fact, the model with rearrangements of $m=4$ particles has $\chi_4^*$ roughly 4 times as large.} 
    \label{fig:app_singlesite}
\end{figure}

As stated in the main text, we decided assume rearrangements involve $m=4$ particles after observing that a model without this effect strongly underestimates the value of $\chi_4^*$. Figure \ref{fig:app_singlesite} compares the $\chi_4^*$ produced by such a model to the values for MD data and our full model, as shown in Figure \ref{fig:chipeak}. 

\section{Scaling collapses to extract \texorpdfstring{$\xi_4$}{the dynamical lengthscale}} \label{app:collapse}

\begin{figure}[h!]
    \includegraphics[]{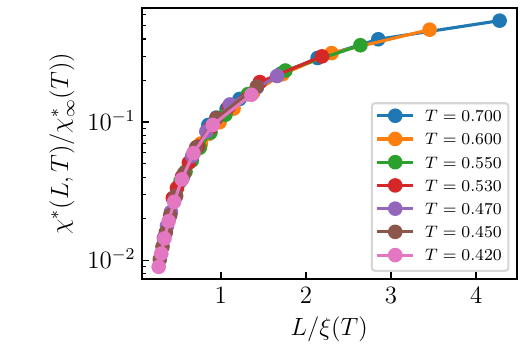}
    \caption{Scaling collapse of block $\chi_4^*$ for MD simulations, used to extract $\xi_4(T)$.}
    \label{fig:MD_xi_collapse}
\end{figure}

\begin{figure}[h!]
    \includegraphics[]{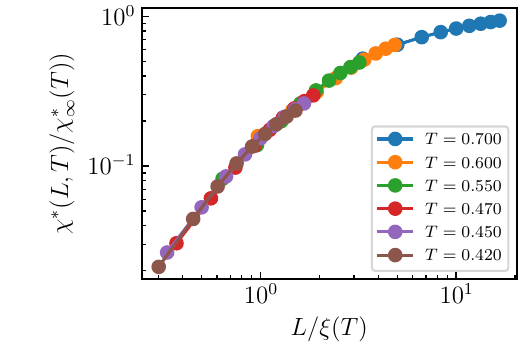}
    \caption{Scaling collapse of block $\chi_4^*$ for TRSP model, used to extract $\xi_4(T)$} 
    \label{fig:model_xi_collapse}
\end{figure}

As in \cite{tah2017block}, we extract the scaling of $\xi_4(T)$ by scaling collapse of $\chi_4^*$ computed for subsystems (``blocks'') of size $L$. These collapses are shown in \ref{fig:MD_xi_collapse} and \ref{fig:model_xi_collapse}.

\begin{figure}[h!]
    \includegraphics[]{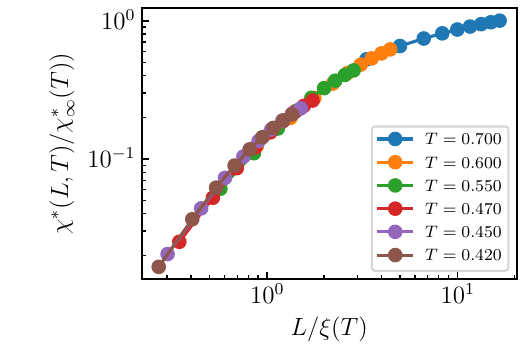}
    \caption{Scaling collapse of block $\chi_4^*$ for model with $c=0.6$.} 
    \label{fig:model_xi_collapse_6}
\end{figure}

\begin{figure}[h!]
    \includegraphics[]{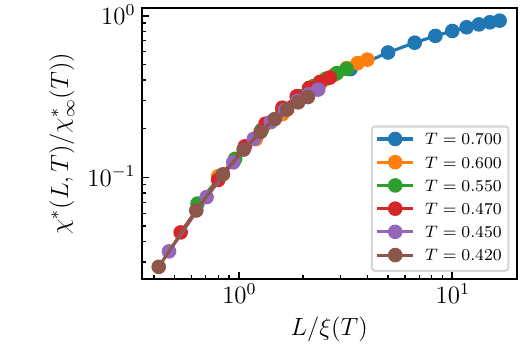}
    \caption{Scaling collapse of block $\chi_4^*$ for model with $c=0.2$.} 
    \label{fig:model_xi_collapse_2}
\end{figure}

\clearpage

\bibliography{Dynamics}

\end{document}